\documentclass[pra, twosides, twocolumn]{revtex4}
\usepackage{amssymb}

\usepackage{graphicx, amsmath, amssymb}

\begin{document}

\title{Total Quantum Zeno Effect beyond Zeno Time }
\author{D. Mundarain$^{1}$, M. Orszag$^{2}$ and J. Stephany$^{1}$}
\address{ \
${}^{1}$ Departmento de F\'{\i}sica, Universidad Sim{\'o}n
Bol{\'\i}var,
Apartado Postal 89000, Caracas 1080A, Venezuela \\
${}^{2}$ Facultad de F\'{\i}sica, Pontificia Universidad
Cat\'{o}lica de Chile, Casilla 306, Santiago, Chile}

\begin{abstract}
In this work we show that is possible to obtain Total Quantum Zeno
Effect in an unstable systems for times larger than the
correlation time of the bath. The effect is observed for some
particular systems  in which one can chose appropriate observables
which frequent measurements freeze the system into the initial
state. For a two level system in a squeezed bath one can show that
there are two bath dependent observables  displaying Total Zeno
Effect when the system is initialized in some particular states.
We show also that  these  states  are intelligent states of two
conjugate observables associated to the electromagnetic
fluctuations of the bath.
\end{abstract}

\maketitle

\section{Introduction}

 An interesting consequence of the fact that frequent measurements  can
 modify the dynamics of  quantum systems is
known as  Quantum Zeno Effect (QZE) \cite{1,2,3,4,5} . In general  QZE
is related to  suppression of induced transitions in interacting systems
or  reduction of the decay rate in unstable systems. Also, the opposite effect of
 enhancing the  decay process by frequent measurements has been  predicted
  and  is known as Anti-Zeno Effect (AZE). The experimental observation of QZE in the early
days was restricted to oscillating quantum systems \cite{6} but recently,
both QZE and AZE were successfully observed in irreversible decaying
processes.\cite{7,8,9}.

 Quantum theory of measurements predicts reduction of the decay rate in
 unstable systems  when the time between successive measurements is smaller
than the Zeno Time which  is known to be  smaller than the correlation time of
the bath. This  effect is universal in the sense that it does not depend on
the measured observable whenever the time between measurements is very
small. This observation does not preclude the manifestation of Zeno
Effect for times larger than the correlation time
for some well selected observables in a particular bath. In this work we show
that  is possible for a  two-level system  interacting
with a squeezed bath  to select a couple observables whose
measurements beyond the correlation time for adequately prepared systems
lead to the total suppression of transitions, i.e Total Zeno Effect.

This work is organized as follows: In section (\ref{sec2}) we discuss some
general facts and review some results obtained in reference \cite{mund}
which are needed for our discussion. In Section (\ref{sec3}) we define the
system we deal with and identify the observables and the corresponding
initial states which are shown to display Total Zeno Effect. In section (\ref
{sec4}) we show that the initial states which show Total Zeno Effect are
intelligent spin states, i.e states that saturate the Heisenberg Uncertainty
Relation for two fictitious spin operators. Finally, we discuss the
results in Section (\ref{sec5}).

\section{Total Zeno effect in unstable systems}

\label{sec2}

Consider a closed system with Hamiltonian $H$ and an observable $A$ with
discrete spectrum. If the initial state of the system is the  eigenstate $%
|a_n\rangle $ of $A$ with eigenvalue $a_n$, the probability of survival in a
sequence of $S$ measurements, that is the probability that in all
measurements one gets the same result $a_n$, is
\begin{equation}
P_n(\Delta t,S) = \left( 1 - \frac{\Delta t^2}{\hbar^2} \Delta_n^2 \mathit{H}%
\right)^S
\end{equation}
where
\begin{equation}
\Delta_n^2 \mathit{H} = \langle a_n |\mathit{H}^2|a_n\rangle -\langle a_n |%
\mathit{H}|a_n\rangle^2
\end{equation}
and $\Delta t$ is the time between consecutive measurements. In the limit of
continuous monitoring ( $S \rightarrow \infty ,\Delta t \rightarrow 0$ and $%
S \Delta t \rightarrow t $ ), $P_n \rightarrow 1$ and the system is freezed
in the initial state.

 In an unstable system and for times larger than the
correlation time of the bath, the irreversible
evolution  of the system can be  described in terms of the Liouville operator $\mathit{L}%
\{\rho \}$ by using the master equation;
\begin{equation}
\frac{\partial \rho }{\partial t}=\mathit{L}\{\rho \}\ .  \label{master}
\end{equation}

In this case  the survival probability in a sequence of $S$ measurements  is:
\begin{equation}
P_n(\Delta t,S) = \left( 1 + \Delta t \quad\langle a_{n}|\mathit{L}\{|a_{n}\rangle \langle
a_{n}|\}|a_{n}\rangle  \right)^S \label{eqn54}
\end{equation}

Then,  the survival probability in the limit of  continuous monitoring  is
time dependent and is easy to show that it is
given by
\begin{equation}
P_{n}(t)=\exp \left\{ \langle a_{n}|\mathit{L}\{|a_{n}\rangle \langle
a_{n}|\}|a_{n}\rangle t\right\} \ \ .  \label{ec13}
\end{equation}

In fact for non zero bath correlation time ($\tau _{D}\neq 0$) one cannot take the continous
monitoring limit and  the equation
 (\ref{ec13}) is an aproximation since $\Delta t$ cannot be strictly zero and
 at the same time be larger than $\tau _{D}$. In that case this expression is
 valid only  when the time between consecutive measurements
is small enough but greater than the correlation time of the bath. For
mathematical simplicity in what follows we consider the zero correlation
time limit  and then one is allowed to take the limit of
continuous monitoring.

>From equation (\ref{ec13}) one observes that the Total
Zeno Effect is possible when
\begin{equation}
\langle a_{n}|\mathit{L}\{|a_{n}\rangle \langle a_{n}|\}|a_{n}\rangle =0\ \ .
\label{ec1239}
\end{equation}
Then, for times larger than the correlation time, the possibility of having
Total Zeno Effect depends on the dynamics of the system ( determined by the
interaction with the baths), on the  observable to be measured and
on the particular eigenstate of the observable chosen as  the  initial
state of the system.

 If equation (\ref{ec1239}) is satisfied,
then equation (\ref{ec13}) must be corrected, taking the next non-zero
contribution in the expansion of $\rho (\Delta t)$. In that case the
eq. (\ref{eqn54}) becomes:
\begin{equation}
P_n(\Delta t,s) = \left( 1 +   \langle a_{n}| \mathit{L} \{
  \mathit{L} \{ |a_{n} \rangle \langle
a_{n}| \} \} |a_{n} \rangle \Delta t^2 / 2 \right)^S
\end{equation}
Then the survival probility for continous monitoring  is
\begin{equation}
P_n(t) = \exp \{  \frac{\langle a_{n}| \mathit{L} \{
  \mathit{L} \{ |a_{n} \rangle \langle
a_{n}| \} \} |a_{n} \rangle \Delta t}{2} \quad t \}
\end{equation}
 In general  $\mathit{L}$ is proportional
to $\gamma$, the decay constant for vacuum. Then as one can see  a decay rate
proportional to $\gamma^{2} \Delta t$ appears. and the decay
time is $\varpropto \frac{1}{\gamma ^{2}\Delta t}$, which is in general a
number much larger than the typical evolution time of the system since $\Delta
t  \ll  \gamma$.
This observation is particularly important for system in which one cannot
take the zero limit in $\Delta t$, i.e when one has a bath with a non zero  correlation time.
 Notice
that as the spectrum of the bath gets broader, $\tau _{D}$ becomes
smaller, and one is able to choose a smaller $\Delta t$, approaching in this
way  the ideal situation and the Total Zeno Effect.

\section{Total Zeno observables}

\label{sec3}

In the interaction picture the Liouville operator for a two level system in
a broadband squeezed vacuum has the following structure \cite{gar},

\begin{eqnarray}
L\{\rho \} &=&\frac{1}{2}\gamma \left( N+1\right) \left( 2\sigma {\rho }%
\sigma^{\dagger}-\sigma^{\dagger} \sigma {\rho }-{\rho }\sigma^{\dagger}\sigma
\right)  \nonumber  \label{em1} \\
&&\frac{1}{2}\gamma N\left( 2\sigma^{\dagger}{\rho }\sigma-\sigma
  \sigma^{\dagger}
{\rho }-{\rho }\sigma \sigma^{\dagger} \right)  \nonumber \\
& &-\gamma M e^{i\phi} \sigma^{\dagger}{\rho} \sigma^{\dagger} - \gamma M
e^{-i\phi } \sigma {\rho} \sigma
\end{eqnarray}
where  $\gamma $ is the vacuum decay constant and $N,M=\sqrt{N(N+1)}$ and
$\psi$
are the parameters of the squeezed bath. Here $\sigma$ and
$\sigma^{\dagger}$ are the  ladder operators for a two level system,

\begin{equation}
\sigma =\frac{1}{2}(\sigma_x-i\sigma_y) \qquad
\sigma^{\dagger}= \frac{1}{2} (\sigma_x+i\sigma_y)
\end{equation}
with $\sigma _{x}$, $\sigma _{y}$ and $\sigma _{z}$ the Pauli matrices.

Let us introduce the Bloch representation of the two level density matrix
\begin{equation}
\rho =\frac{1}{2}\left( 1+\vec{\rho} \cdot \vec{\sigma}\right)  \label{dm1}
\end{equation}

Using this representation and the master equation one can obtain the following
set of differential equation for the components of the Bloch vector $(\rho_x,\rho_y,\rho_z)$:

\begin{eqnarray}
\dot{\rho _{x}} &=&- \gamma \left( N +1/2 + M \cos(\psi) \right) \rho _{x}
+\gamma M \sin (\psi)\rho _{y} \nonumber \\
\dot{\rho _{y}} &=&- \gamma \left( N +1/2 -M \cos(\psi) \right) \rho _{y}
+\gamma M \sin (\psi)\rho _{x} \nonumber \\
\dot{\rho _{z}} &=&- \gamma \left( 2 N +1 \right) \rho _{z} -\gamma
\label{ec1222}
\end{eqnarray}

which has the following solutions:
\begin{eqnarray}
\rho _{x}(t) &=&\left( \rho _{x}(0)\sin ^{2}(\psi /2)+\rho _{y}(0)\sin (\psi
/2)\cos (\psi /2)\right)  \nonumber \\
&&e^{-\gamma (N+1/2-M)\,t}  \nonumber \\
&&+\left( \rho _{x}(0)\cos ^{2}(\psi /2)-\rho _{y}(0)\sin (\psi /2)\cos
(\psi /2)\right)  \nonumber \\
&&e^{-\gamma (N+1/2+M)\,t}  \label{ec1222}
\end{eqnarray}
\begin{eqnarray}
\rho _{y}(t) &=&\left( \rho _{y}(0)\cos ^{2}(\psi /2)+\rho _{x}(0)\sin (\psi
/2)\cos (\psi /2)\right)  \nonumber \\
&&e^{-\gamma (N+1/2-M)\,t}  \nonumber \\
&&+\left( \rho _{y}(0)\sin ^{2}(\psi /2)-\rho _{x}(0)\sin (\phi /2)\cos
(\psi /2)\right)  \nonumber \\
&&e^{-\gamma (N+1/2+M)\,t}  \label{ec1223}
\end{eqnarray}
\begin{equation}
\rho _{z}(t)=\rho _{z}(0)e^{-\gamma (2N+1)t}+\frac{1}{2N+1}\left( e^{-\gamma
(2N+1)t}-1\right)
\end{equation}

These equations describe  the
behavior of the system when there are no measurements.

 Consider  now the hermitian operator $\sigma _{\mu }$ associated to the
fictitious spin component in the direction of the unitary vector $\hat{\mu}%
=(\cos (\phi )\sin (\theta ),\sin (\phi )\sin (\theta ),\cos (\theta ))$
defined by the angles $\theta $ and $\phi $,
\begin{equation}
\sigma _{\mu }=\vec{\sigma}\cdot \hat{\mu}=\sigma _{x}\cos (\phi )\sin
(\theta )+\sigma _{y}\sin (\phi )\sin (\theta )+\sigma _{z}\cos (\theta )
\end{equation}
The eigenstates of $\sigma _{\mu }$ are,
\begin{equation}
|+\rangle _{\mu }=\cos (\theta /2)\,|+\rangle +\sin (\theta /2)\,\exp {%
(i\phi )}\,|-\rangle
\end{equation}
\begin{equation}
|-\rangle _{\mu }=-\sin (\theta /2)\,|+\rangle +\cos (\theta /2)\,\exp {%
(i\phi )}\,|-\rangle
\end{equation}

If the system is initialized in the state $|+\rangle _{\mu }$ the survival
probability at time $t$ is
\begin{equation}
P_{\mu }^{+}(t)=\exp \left\{ F(\theta ,\phi )\,\,t\,\right\}
\end{equation}
where
\begin{equation}
F(\theta ,\phi )=\,_{\mu }\langle +|\,\,L\left\{ \,\,|+\rangle _{\mu
}\,_{\mu }\langle +|\,\,\right\} \,\,|+\rangle _{\mu }\ \ .
\end{equation}
In this case the function $F(\theta ,\phi )$ has the structure
\begin{eqnarray}
F(\theta ,\phi ) &=&-\frac{1}{2}\gamma \left( N+1\right) \left( \rho
_{z}(0)+\rho _{z}^{2}(0)+\frac{1}{2}\rho _{x}^{2}(0)+\frac{1}{2}{\rho
_{y}^{2}(0)}\right)  \nonumber \\
&&+\frac{1}{2}\gamma N\left( (\rho _{z}(0)-\rho _{z}^{2}(0)-\frac{1}{2}\rho
_{x}^{2}(0)-\frac{1}{2}{\rho _{y}^{2}(0)}\right)  \nonumber \\
&&-\frac{1}{2}\gamma M\rho _{x}(0)(\cos (\psi )\rho _{x}(0)-\sin (\psi ){%
\rho _{y}(0)})  \nonumber \\
&&+\frac{1}{2}\gamma M\rho _{y}(0)(\sin (\psi )\rho _{x}(0)+\cos (\psi ){%
\rho _{y}(0)})
\end{eqnarray}
where now $\vec{\rho}(0)=\hat{\mu}$ is a function of the angles..

\begin{figure}[h]
\includegraphics[width=6cm,angle=-90]{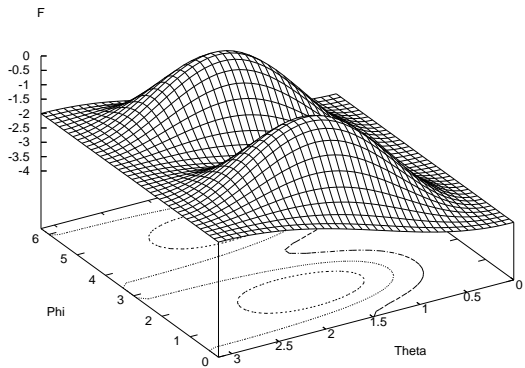}\bigskip
\caption{$F(\phi,\theta)$ for $N=1$ and $\psi =0$ }
\label{etiqueta1}
\end{figure}

In figure (\ref{etiqueta1}) we show $F(\phi ,\theta )$ for $N=1$ and $\psi
=0 $ as function of $\phi $ and $\theta $. The maxima correspond to $F(\phi
,\theta )=0$. For arbitrary values of $N$ and $\psi$ there are two maxima
corresponding to the following angles:

\begin{equation}
\phi_1^M = \frac{\pi-\psi}{2} \quad \mathrm{and} \quad \cos(\theta^M) = -
\frac{1}{2 \left( N+M+ 1/2\right)}
\end{equation}
and
\begin{equation}
\phi_2^M = \frac{\pi-\psi}{2}+\pi \quad \mathrm{and} \quad \cos(\theta^M) =
- \frac{1}{2 \left( N+M+ 1/2\right)}
\end{equation}

These preferential directions given by the vectors $\mathbf{\hat{\mu}_1}
=(\cos(\phi_1^M)\sin(\theta^M),\sin(\phi_1^M)\sin (\theta^M),\cos(\theta^M))$
and $\mathbf{\hat{\mu}_2}=(\cos(\phi_2^M)\sin(\theta^M),\sin(\phi_2^M)\sin(
\theta^M),\cos(\theta^M))$) define the operators $\sigma _{\mu_1}$ and $
\sigma _{\mu_2}$ which show Total Zeno Effect if the initial state of the
system is the eigenstate $|+\rangle _{\mu_1}$ or respectively
$|+\rangle_{\mu_2}$, then   each preferential  observable has only one
eigenstate displaying Total Zeno Effect.  These eigenstates are:

\begin{equation}
|+\rangle _{\mu_1} = \sqrt{\frac{N}{N+M}}|+\rangle +i \sqrt{\frac{M}{N+M}}
 \exp \{- i \frac{\psi}{2} \} |-\rangle
\end{equation}
and
\begin{equation}
|+\rangle _{\mu_2} = \sqrt{\frac{N}{N+M}}|+\rangle -i \sqrt{\frac{M}{N+M}}
 \exp \{- i \frac{\psi}{2} \} |-\rangle
\end{equation}

The other eigenstates of the observables do not display Total Zeno Effect.
 As final remark is important to observe that in the previous
calculations  we  have ever chosen the state $|+\rangle _{\mu}$
in order to optimize the function $F(\phi ,\theta) $. In fact one can select
the state  $|-\rangle _{\mu}$ but the final observables
displaying Total Zeno Effect will be in the same preferential directions
indicated above.

\section{Master equation and measurements}

Besides of the Total Zeno effect obtained in the cases specified previously it
is also very interesting to discuss the effect of measurements for other
choices of the initial state, the states which do not display Total Zeno
Effect. To be specific let us consider measurements of the observable
$\sigma_{\mu}= \vec{\sigma} \cdot \hat{\mu}$. The modified master equation with
the measurement of $\sigma_{\mu}$ is given by \cite{mund}:

\begin{equation}
\frac{\partial \rho }{\partial t}=P_{\mu}\,L\left\{ \rho \right\}
\,P_{\mu}+\left( 1-P_{\mu}\right) \,L\left\{ \rho \right\}
\,\left( 1-P_{\mu}\right)\label{ecmcm}
\end{equation}

where

\begin{equation}
P_{\mu}=|+\rangle _{\mu}\,_{\mu }\langle +|
\end{equation}

and $\mathit{L}\{\rho \}$ is given by (\ref{em1}).
This equation can be solved
using the Bloch representation of the density  matrix. In this case we  can
write the density operator in terms of   a second set of rotated Pauli matrices  that  includes  the Pauli observable which we are measuring :

\begin{equation}
\rho =\frac{1}{2}\left( 1+\rho_{\mu} \sigma_{\mu}+\rho_{\alpha} \sigma_{\alpha} +\rho_{\beta} \sigma_{\beta}\right)  \label{dm1}
\end{equation}
where $\sigma_{\alpha}$ and $\sigma_{\beta}$ are two Pauli matrices projected
in two orthogonal direction to the vector $\hat{\mu}$. During the process of
measurement one  obtains always eigenvectors of
$\sigma_{\mu}$ observable, these eigenvectors have the property of being zero
valued for the other two observables. Then during the measurement process the
quantities $\rho_{\alpha}$  and $\rho_{\beta}$ are equal to zero because these
quantities coprrespond to the mean values of the respectives observables. Then
in this case the density matrix can be written in term of one parameter which
corresponds to the mean value of the observable that is bein measured;
\begin{equation}
\rho =\frac{1}{2}\left( 1+\rho_{\mu} \sigma_{\mu}\right)  \label{dm1}
\end{equation}
with
\begin{equation}
\rho_{\mu} = \langle  \sigma_{\mu}\rangle = {\rm Tr}\left\{ \rho \sigma_{\mu}
\right\}
\end{equation}
Then the master equation is reduced to the following differential equation:
\begin{displaymath}
\dot{\rho_{\mu}} = {\rm Tr}\left\{ \dot{\rho} \sigma_{\mu}\right\}
\end{displaymath}
\begin{displaymath}
  ={\rm Tr} \left\{ \left(P_{\mu}\,L\left\{ \rho \right\}
\,P_{\mu}+\left( 1-P_{\mu}\right) \,L\left\{ \rho \right\}
\,\left( 1-P_{\mu}\right) \right) \sigma_{\mu}\right\} \end{displaymath}
\begin{equation}
={\rm Tr} \left\{ L\left\{ \rho \right\} \sigma_{\mu}\right\} \label{eq145}
\end{equation}
This equation could induced to think that the evolution with and
withouth measurements are equal, but we must remember that the
density matrix in the right hand side of (\ref{ec145}) is the
density with measurements.  Substituting the form of the density
matrix during the measuring process one can obtain a real
differential equation for $\rho_{\mu}$:
\begin{equation}
\dot{\rho_{\mu}} =\alpha+ \beta \rho_{\mu}
\end{equation}
where
\begin{equation}
\alpha =\frac{1}{2}  {\rm Tr}\left\{\,L\left\{ 1 \right\}
  \sigma_{\mu}\right\}
\end{equation}
\begin{equation}
\beta =\frac{1}{2}  {\rm Tr}\left\{\,L\left\{ \sigma_{\mu} \right\}
  \sigma_{\mu}\right\}
\end{equation}
In our case and measuring $\sigma_{\mu_1}$ one obtains
\begin{equation}
\alpha = 2 \, \gamma \,\left( N-M+1/2\right)
\end{equation}
and
\begin{equation}
\beta = -\alpha = -2 \, \gamma \, \left( N-M+1/2\right)
\end{equation}
The solution to the differential equation is
\begin{equation}
\rho_{\mu}(t)  =  1+ \left( \rho_{\mu}(o) -1\right) e^{-\alpha t}
\end{equation}
one can observe the Total Zeno Effect when  $\rho_{\mu}(o)=1$ which correspond to having as  initial
state  $|+\rangle _{\mu _{1}}$.

\begin{figure}[h]
\includegraphics[width=6cm,angle=-90]{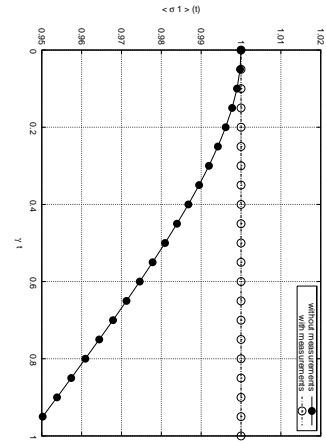}\bigskip
\caption{$\langle \sigma _{\mu _{1}}(t)\rangle $ for $N=1$ and $\psi =0$.
Solid circles: no measurements. Empty circles: with measurements. One
measures $\sigma _{\mu _{1}}$ and the initial state is $|+\rangle _{\mu
_{1}} $ }
\label{etiqueta2}
\end{figure}

In figure (\ref
{etiqueta2}) we show the evolution of $\langle \sigma _{\mu _{1}}\rangle $,
that is the mean value of observable $\sigma _{\mu _{1}}$, when the system
is initialized in the state $|+\rangle _{\mu _{1}}$.  without measurements
(master equation (\ref{em1})) and with frequent monitoring of $\sigma _{\mu
_{1}}$ (master equation (\ref{ecmcm})). Consistently with our discussion of
frequent measurements, the system is freezed in the state $|+\rangle _{\mu
_{1}}$ (Total Zeno Effect).

\begin{figure}[h]
\includegraphics[width=6cm,angle=-90]{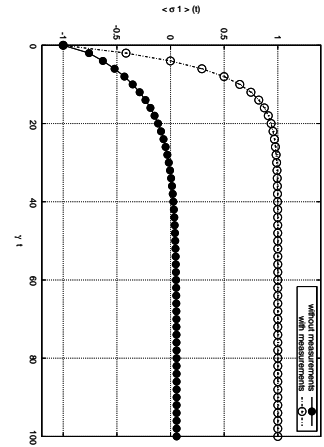}\bigskip
\caption{$\langle \sigma _{\mu _{1}}(t)\rangle $ for $N=1$ y $\psi =0$.
Solid circles: no measurements. Empty circles: with measurements. One
measures $\sigma _{\mu _{1}}$ and the initial state is $|-\rangle _{\mu
_{1}} $}
\label{etiqueta3}
\end{figure}

In figure (\ref{etiqueta3}) we show the time evolution of $\langle \sigma
_{\mu _{1}}\rangle $ when the initial state is $|-\rangle _{\mu _{1}}$
without measurements and with measurements of the same observable as in
previous case. One observes that with measurements the system evolves from $%
|-\rangle _{\mu _{1}}$ to $|+\rangle _{\mu _{1}}$. In general for any
initial state the system under frequent measurements evolves to $|+\rangle
_{\mu _{1}}$ which is the stationary state of Eq. ( \ref{ecmcm}) whenever we
do measurements in $\sigma _{\mu _{1}}$. Analogous effects are observed if
one measures $\sigma _{\mu _{2}}$. In contrast, for measurements in other
directions different from those defined by $\mathbf{\hat{\mu}_{1}}$ or $%
\mathbf{\hat{\mu}_{2}}$ , the system evolves to states which are not
eigenstates of the measured observables.

\section{Intelligent States}

\label{sec4}

Aragone et al \cite{ar} considered well defined angular momentum states that
satisfy the equality $(\Delta J_{x}\Delta J_{y})^{2}=\frac{1}{4}\mid \langle
J_{z}\rangle \mid ^{2}$ in the uncertainty relation. They are called
Intelligent States in the literature. The difference with the coherent or
squeezed states, associated to harmonic oscillators, is that these
Intelligent States are not Minimum Uncertainty States (MUS), since the
uncertainty is a function of the state itself.

In this section we show that the states $|+\rangle _{\mu _{1}}$ and $%
|+\rangle _{\mu _{2}}$ are intelligent states of two observables associated
to the bath fluctuations. The master equation (\ref{em1}) can be written in
an explicit Lindblad form
\begin{equation}
\frac{\partial \rho }{\partial t}=\frac{\gamma }{2}\left\{ 2S\rho S^{\dagger
}-\rho S^{\dagger }S-S^{\dagger }S\rho \right\}  \label{em2}
\end{equation}
using only one Lindblad operator $S$,
\begin{equation}
S = \sqrt{N+1}\sigma -\sqrt{N}\exp \left\{ i\psi \right\} \sigma ^{\dagger}
\end{equation}
\begin{equation}
S = \cosh (r) \sigma -\sinh (r) \exp \left\{ i\psi \right\} \sigma ^{\dagger}
\end{equation}

Obviously any eigenstate of $S$ satisfies the condition (\ref{ec1239}).
It is  very easy to show that the $S$ operator has two eigenvectors
 $|\lambda _{\pm }\rangle$
with eigenvalues $\lambda _{\pm }=\pm i\sqrt{M}\exp \{i\psi /2\}$. It is also
easy  to observe that these two states  are exactly the same states founded in
the previous section,  $|\lambda_+\rangle=|+\rangle_{\mu_1}$ and
$|\lambda_-\rangle=|+\rangle_{\mu_2}$.

Consider now the standard fictitious angular momentum operators for the two
level system are $\{J_{x}=\sigma _{x}/2,J_{y}=\sigma _{y}/2,J_{z}=\sigma
_{z}/2\}$ and also two rotated operators $J_{1}$ and $J_{2}$ which are
consistent with the electromagnetic bath fluctuations in phase space (see
fig. 2 in ref \cite{mund}) and which satisfy the same Heisenberg uncertainty
relation that $J_{x}$ and $J_{y}$ . They are,
\begin{eqnarray}
J_{1} &=&\exp \{i\psi /2J_{z}\}J_{x}\exp \{-i\psi /2J_{z}\}  \nonumber \\
&=&\cos (\psi /2)J_{x}-\sin (\psi /2)J_{y}
\end{eqnarray}
\begin{eqnarray}
J_{2} &=&\exp \{i\psi /2J_{z}\}J_{y}\exp \{-i\psi /2J_{z}\}  \nonumber \\
&=&\sin (\psi /2)J_{x}+\cos (\psi /2)J_{y}
\end{eqnarray}
These two operators are associated respectively with the major and minor
axes of the ellipse which represents the fluctuations of bath.

In terms of $J_{1}$ y $J_{2}$ we have
\begin{equation}
J_{-}=\sigma =(J_{x}-iJ_{y})=\exp \{i\psi /2\}(J_{1}-iJ_{2})\ ,
\end{equation}
\begin{equation}
J_{+}=\sigma ^{\dagger }=(J_{x}+iJ_{y})=\exp \{-i\psi /2\}(J_{1}+iJ_{2})\ .
\end{equation}
Then $S$ can be written in the following  form:
\begin{equation}
S=\exp \{i\psi /2\}\left( \cosh (r)-\sinh (r)\right) \left( J_{1}-i\alpha
J_{2}\right)
\end{equation}
with
\begin{equation}
\alpha =\frac{\cosh (r)+\sinh (r)}{\cosh (r)-\sinh (r)}=\exp \{2r\}
\end{equation}

Following Rashid \textit{et al (}\cite{ra}\textit{)} we define a non
hermitian operator $J_{-}(\alpha )$
\begin{equation}
J_{-}(\alpha )=\frac{\left( J_{1}-i\alpha J_{2}\right) }{(1-\alpha
^{2})^{1/2}}
\end{equation}
so that
\begin{equation}
S=\exp \{i\psi /2\}\left( \cosh (r)-\sinh (r)\right) \,(1-\alpha
^{2})^{1/2}\,J_{-}(\alpha )
\end{equation}
After some algebra one obtains  that
\begin{equation}
S=2\,\lambda _{+}\,J_{-}(\alpha )
\end{equation}

>From this equation one can observe that the eigenstates of $S$ are also  eigenstates of $J_{-}(\alpha )$ with
eigenvalues $\pm 1/2$. It is known that the eigenstates of $J_{-}(\alpha )$ are
 intelligent states of  $J_{1}$ and $J_{2}$, \textit{i.e} they satisfy the equality
condition in the Heisenberg uncertainty relation for these observables:
\begin{equation}
\Delta ^{2}J_{1}\Delta ^{2}J_{2}=\frac{|\langle J_{z}\rangle |^{2}}{4}
\end{equation}

\section{Discussion}

\label{sec5}

We have shown that Total Zeno Effect is obtained for two particular
observables $\sigma _{\mu_1}$ or $\sigma _{\mu_2}$, for which the azimuthal
phases in the fictitious spin representation depend on the phase of the
squeezing parameter of the bath and the polar phases depend on the squeeze
amplitude. In this sense, the parameters of the squeezed bath specify
some definite atomic directions.

When performing frequent measurements on $\sigma _{\mu _{1}}$, starting from
the initial state $|+\rangle _{\mu _{1}}$, the system freezes at the initial
state as opposed to the usual decay when no measurements are done. On the
other hand, if the system is initially prepared in the state $|-\rangle
_{\mu _{1}},$ the frequent measurements on $\sigma _{\mu _{1}}$ will makes
it evolve from the state $|-\rangle _{\mu _{1}}$ to $|+\rangle _{\mu _{1}}$.
More generally, when performing the measurements on $\sigma _{\mu _{1}}$,
any initial state evolves to the same state $|+\rangle _{\mu _{1}}$ which is
the steady state of the master equation (\ref{ecmcm}) in this situation.

The above discussion could appear at a first sight surprising. However,
taking a more familiar case of a two-level atom in contact with a thermal
bath at zero temperature, if one starts from any initial state, the atom
will necessarily decay to the ground state. This is because the time
evolution of $\langle \sigma _{z}\rangle $ is the same with or without
measurements of $\sigma _{z}$. In both cases the system goes to the ground
state, which is an eigenstate of the measured observable $\sigma _{z}$. In
the limit $N,M\rightarrow 0$, $\sigma _{\mu _{1}}\rightarrow -\sigma _{z}$,
and the state $|+\rangle _{\mu _{1}}\rightarrow $ $|-\rangle _{z}$, which
agrees with the known results.

Finally, we  found that the two  eigentates of the two preferential observables
displaying  QZE are also eigenstates of $S$ operator and consequently  intelligent states
of $J_{1},J_{2}$ which are  rotated versions of  $J_{x},J_{y}$ obsevables.

\subsection{Acknowledgements}

Two of the authors(D.M. and J.S.) were supported by Did-Usb Grant Gid-30 and
by Fonacit Grant No G-2001000712.

M.O was supported by Fondecyt \# 1051062 and Nucleo Milenio ICM(P02-049)

\bigskip

\end{document}